\documentclass[twocolumn,10pt]{IEEEtran}

\usepackage{amsmath,epsfig}
\usepackage{amssymb}
\usepackage{amsthm}
\usepackage{color}
\usepackage{url}
\usepackage{cite}
\usepackage{subcaption}

\usepackage{algorithmicx}
\usepackage{algpseudocode}

\usepackage{import}

\newcommand{\maximize}[1]{{\underset{{#1}}{\mathrm{maximize}}}}

\theoremstyle{remark}

\begin{document}

\title{Massive MIMO: \\ Ten Myths and One Critical Question}

\author{Emil Bj{\"o}rnson, Erik G.~Larsson, and Thomas L.~Marzetta}

\maketitle

\begin{abstract}
Wireless communications is one of the most successful technologies in modern years, given that an exponential growth rate in wireless traffic has been sustained for over a century (known as Cooper's law). This trend will certainly continue driven by new innovative applications; for example, augmented reality and internet-of-things.

Massive MIMO (multiple-input multiple-output) has been identified as a key technology to handle orders of magnitude more data traffic. Despite the attention it is receiving from the communication community, we have personally witnessed that Massive MIMO is subject to several widespread misunderstandings, as epitomized by following (fictional) abstract:

\emph{``The Massive MIMO technology uses a nearly infinite number of high-quality antennas at the base stations. By having at least an order of magnitude more antennas than active terminals, one can exploit asymptotic behaviors that some special kinds of wireless channels have. This technology looks great at first sight, but unfortunately the signal processing complexity is off the charts and the antenna arrays would be so huge that it can only be implemented in millimeter wave bands.''}

The statements above are, in fact, completely false. In this overview article, we identify ten myths and explain why they are not true. We also ask a question that is critical for the practical adoption of the technology and which will require intense future research activities to answer properly. We provide references to key technical papers that support our claims, while a further list of related overview and technical papers can be found at the Massive MIMO Info Point: http://massivemimo.eu

\end{abstract}

\section{Introduction}

\begin{figure*}
        \centering
        \begin{subfigure}[b]{1.7\columnwidth} \centering 
                \includegraphics[width=.99\textwidth]{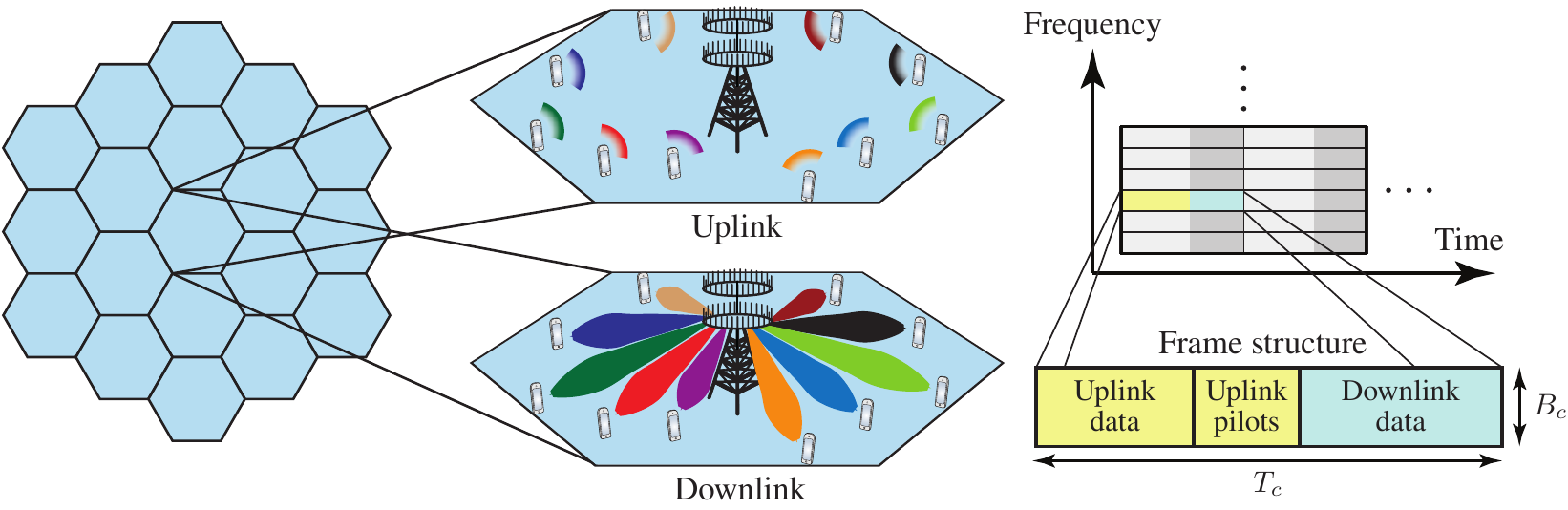} 
                \caption{}
                \label{figure_uplink-downlink}
        \end{subfigure} \\  \vspace{2mm}
        \begin{subfigure}[b]{1.7\columnwidth} \centering
                \includegraphics[width=.6\columnwidth]{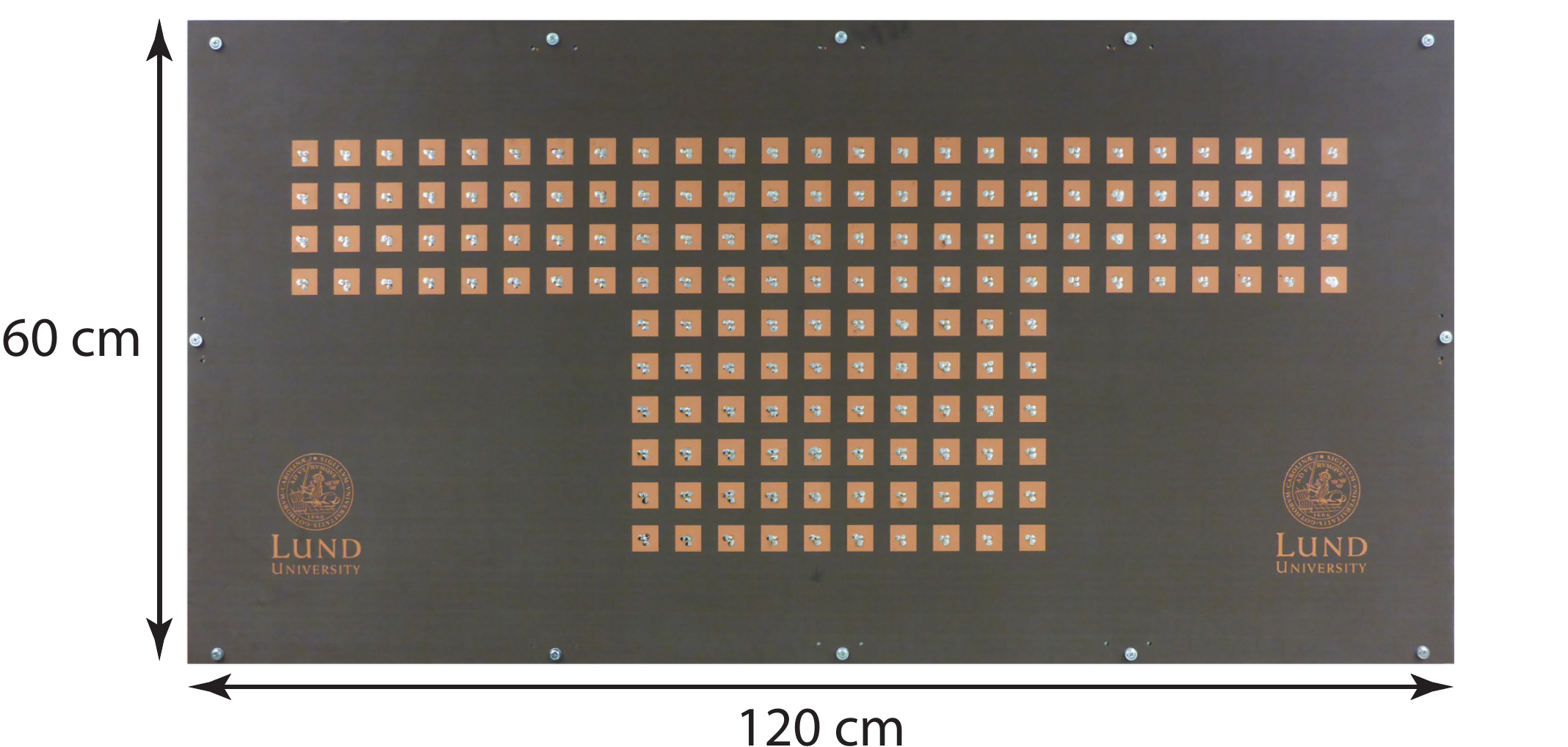} 
                \caption{}
                \label{fig:myth3}
        \end{subfigure}
        \caption{Example of a Massive MIMO system. (a) Illustration of the uplink and downlink in line-of-sight propagation, where each BS is equipped with $M$ antennas and serves $K$ terminals. The TDD transmission frame consists of $\tau = B_c T_c$ symbols. By capitalizing on channel reciprocity, there is payload data transmission in both the uplink and downlink, but only pilot transmission in the uplink. (b) Photo of the antenna array of the LuMaMi testbed at Lund University in Sweden \cite{Vieira2014a}. The array consists of 160 dual-polarized patch antennas. It is designed for a carrier frequency of 3.7 GHz and the element-spacing is 4 cm (half a wavelength).}\label{fig:protocolAndLuMaMi}
\end{figure*}

Massive MIMO is a multi-user MIMO technology where each base station (BS) is equipped with an array of $M$ active antenna elements and utilizes these to communicate with $K$ single-antenna terminals---over the same time and frequency band. The general multi-user MIMO concept has been around for decades, but the vision of actually deploying BSs with more than a handful of service antennas is relatively new \cite{Marzetta2010a}. By coherent processing of the signals over the array, transmit precoding can be used in the downlink to focus each signal at its desired terminal and receive combining can be used in the uplink to discriminate between signals sent from different terminals. The more antennas that are used, the finer the spatial focusing can be. An illustration of these concepts is given in Figure \ref{figure_uplink-downlink}.

The canonical Massive MIMO system operates in time-division duplex (TDD) mode, where the uplink and downlink transmissions take place in the same frequency resource but are separated in time. The physical propagation channels are reciprocal---meaning that the channel responses are the same in both directions---which can be utilized in TDD operation. In particular, Massive MIMO systems exploit the reciprocity to estimate the channel responses on the uplink and then use the acquired channel state information (CSI) for both uplink receive combining and downlink transmit precoding of payload data. Since the transceiver hardware is generally not reciprocal, calibration is needed to exploit the channel reciprocity in practice. Fortunately, the uplink-downlink hardware mismatches only change by a few degrees over a one-hour period and can be mitigated by simple relative calibration methods, even without extra reference transceivers and by only relying on mutual coupling between antennas in the array \cite{Vieira2014a}.

There are several good reasons to operate in TDD mode. Firstly, only the BS needs to know the channels to process the antennas coherently. Secondly, the uplink estimation overhead is proportional to the number of terminals, but independent of $M$ thus making the protocol fully scalable with respect to the number of service antennas. Furthermore, basic estimation theory tells us that the estimation quality (per antenna) cannot be reduced by adding more antennas at the BS---in fact, the estimation quality improves with $M$ if there is a known correlation structure between the channel responses over the array \cite{Yin2013a}. 

Since fading makes the channel responses vary over time and frequency, the estimation and payload transmission must fit into a time/frequency block where the channels are approximately static. The dimensions of this block are essentially given by the coherence bandwidth $B_c$ Hz and the coherence time $T_c$ s,  which fit $\tau = B_c T_c$ transmission symbols. Massive MIMO can be implemented either using single-carrier or multi-carrier modulation. We consider multi-carrier OFDM modulation here for simplicity, because the coherence interval has a neat interpretation: it spans a number of subcarriers over which the channel frequency response is constant, and a number of OFDM symbols over which the channel is constant; see Figure~\ref{figure_uplink-downlink}. The channel coherency depends on the propagation environment, user mobility, and the carrier frequency.

\subsection{Linear Processing}

The payload transmission in Massive MIMO is based on linear processing at the BS. In the uplink, the BS has $M$ observations of the multiple access channel from the $K$ terminals. The BS applies linear receive combining to discriminate the signal transmitted by each terminal from the interfering signals. The simplest choice is maximum ratio (MR) combining that uses the channel estimate of a terminal to maximize the strength of that terminal's signal, by adding the signal components coherently. This results in a signal amplification proportional to $M$, which is known as an array gain. Alternative choices are zero-forcing (ZF) combining, which suppresses inter-cell interference at the cost of reducing the array gain to $M-K+1$, and minimum mean squared error (MMSE) combining that balances between amplifying signals and suppressing interference.

The receive combining creates one effective scalar channel per terminal where the intended signal is amplified and/or the interference is suppressed. Any judicious receive combining will improve by adding more BS antennas, since there are more channel observations to utilize. The remaining interference is typically treated as extra additive noise, thus conventional single-user detection algorithms can be applied. Another benefit from the combining is that small-scale fading averages out over the array, in the sense that its variance decreases with $M$. This is known as \emph{channel hardening} and is a consequence of the law of large numbers.

Since the uplink and downlink channels are reciprocal in TDD systems, there is a strong connection between receive combining in the uplink and the transmit precoding in the downlink \cite{Bjornson2016a}---this is known as uplink-downlink duality. Linear precoding based on MR, ZF, or MMSE principles can be applied to focus each signal at its desired terminal (and possibly mitigate interference towards other terminals).

Many convenient closed-form expressions for the achievable uplink or downlink spectral efficiency (per cell) can be found in the literature; see \cite{Ngo2013a,Hoydis2013a,Bjornson2016a} and references therein. We provide an example for i.i.d.~Rayleigh fading channels with MR processing, just to show how beautifully simple these expressions are:
\begin{align} \label{eq:uplink-capacity}
K \cdot \left( 1 - \frac{K}{\tau} \right) \cdot \log_2 \left( 1+ \frac{ c_{ \textrm{\tiny CSI}} \cdot M \cdot \mathrm{SNR}_{u/d}}{K \cdot \mathrm{SNR}_{u/d} + 1} \right) \quad \textrm{[bit/s/Hz/cell]}.
\end{align}
where $K$ is the number of terminals, $( 1 - \frac{K}{\tau} )$ is the loss from pilot signaling, and $\mathrm{SNR}_{u/d}$ equals the uplink signal-to-noise ratio (SNR), $\mathrm{SNR}_{u}$, when Eq.~\eqref{eq:uplink-capacity} is used to compute the uplink performance. Similarly, we let $\mathrm{SNR}_{u/d}$ be the downlink SNR, $\mathrm{SNR}_{d}$, when Eq.~\eqref{eq:uplink-capacity} is used to measure the downlink performance. In both cases, $c_{\textrm{\tiny CSI}} = ({1} + \frac{1}{K \cdot \mathrm{SNR}_u})^{-1}$ is the quality of the estimated CSI, proportional to the mean-squared power of the MMSE channel estimate (where $c_{\textrm{\tiny CSI}} =1$ represents perfect CSI). Notice how the numerator inside the logarithm increases proportionally to $M$ due to the array gain and that the denominator represents the interference plus noise.

While canonical Massive MIMO systems operate with single-antenna terminals, the technology also handles $N$-antenna terminals. In this case, $K$ denotes the number of simultaneous data streams and \eqref{eq:uplink-capacity} describes the spectral efficiency per stream. These streams can be divided over anything from $K/N$ to $K$ terminals, but we focus on $N=1$ in this paper for clarity in presentation.

\section*{Myths and Misunderstandings About Massive MIMO}

The interest in the Massive MIMO technology has grown quickly in recent years, but at the same time we have noticed that there are several widespread myths or misunderstandings around its basic characteristics. This article inspects ten common beliefs concerning Massive MIMO and explains why they are erroneous.

\subsection*{\small Myth 1: Massive MIMO is only suitable for millimeter wave bands}

Antenna arrays are typically designed with an antenna spacing of at least $\lambda_c/2$, where $\lambda_c$ is the wavelength at the intended carrier frequency $f_c$. Larger antenna spacings provide less correlated channel responses over the antennas and thus more spatial diversity, but the important thing in Massive MIMO is that each terminal has distinct spatial channel characteristics and not that the antennas observe uncorrelated channels. The wavelength is inversely proportional to $f_c$, thus smaller form factors are possible at higher frequencies (e.g., in millimeter bands). Nevertheless, Massive MIMO arrays have realistic form factors also at a typical cellular frequency of $f_c=2$ GHz; the wavelength is $\lambda_c = 15$ cm and up to 400 dual-polarized antennas can thus be deployed in a  $1.5 \times 1.5$ meter array. This should be compared with contemporary cellular networks that utilize vertical panels, around 1.5 meter tall and 20 cm wide, each comprising many interconnected radiating elements that provide a fixed directional beam. A 4-MIMO setup uses four such panels with a combined area comparable to the exemplified Massive MIMO array.

\textbf{Example:} Figure \ref{fig:myth3} shows a picture of the array in the LuMaMi Massive MIMO testbed \cite{Vieira2014a}. It is designed for a carrier frequency of $f_c=3.7$ GHz, which gives $\lambda_c=8.1$ cm. The panel is $60 \times 120$ cm (i.e., equivalent to a 53 inch flat-screen TV) and features 160 dual-polarized antennas, while leaving plenty of room for additional antenna elements. Such a panel could easily be deployed at the facade of a building.

The research on Massive MIMO has thus far focused on cellular frequencies below 6 GHz, where the transceiver hardware is very mature. The same concept can definitely be applied in millimeter wave bands as well---many antennas might even be required in these bands since the effective area of an antenna is much smaller. However, the hardware implementation will probably be quite different from what has been considered in the Massive MIMO literature; see, for example, \cite{Alkhateeb2014a}. Moreover, for the same mobility the coherence time will be an order of magnitude shorter due to higher Doppler spread \cite{Adhikary2014a}, which reduces the spatial multiplexing capability. In summary, Massive MIMO for cellular bands and for millimeter bands are two feasible branches of the same tree, whereof the former is mature and the latter is greatly unexplored and possesses many exciting research opportunities.

\subsection*{\small Myth 2: Massive MIMO only works in rich-scattering environments}

The channel response between a terminal and the BS can be represented by an $M$-dimensional vector. Since the $K$ channel vectors are mutually non-orthogonal in general, advanced signal processing (e.g., dirty paper coding) is needed to suppress interference and achieve the sum capacity of the multi-user channel. \emph{Favorable propagation} (FP) denotes an environment where the $K$ users' channel vectors are mutually orthogonal (i.e., their inner products are zero). FP channels are ideal for multi-user transmission since the interference is removed by simple linear processing (i.e., MR and ZF) that utilizes the channel orthogonality \cite{ngo2014aspects}. The question is whether there exist any FP channels in practice?

An approximate form of favorable propagation is achieved in non-line-of-sight (non-LoS) environments with rich scattering, where each channel vector has independent stochastic entries with zero mean and identical distribution. Under these conditions, the inner products (normalized by $M$) go to zero as more antennas are added; this means that the channel vectors get closer and closer to orthogonal as $M$ increases. The sufficient condition above is satisfied for Rayleigh fading channels, which are considered in the vast majority of works on Massive MIMO, but approximate favorable propagation is obtained in many other situations as well.

\begin{figure}
        \centering
        \begin{subfigure}[b]{\columnwidth} \centering 
                \includegraphics[width=\columnwidth]{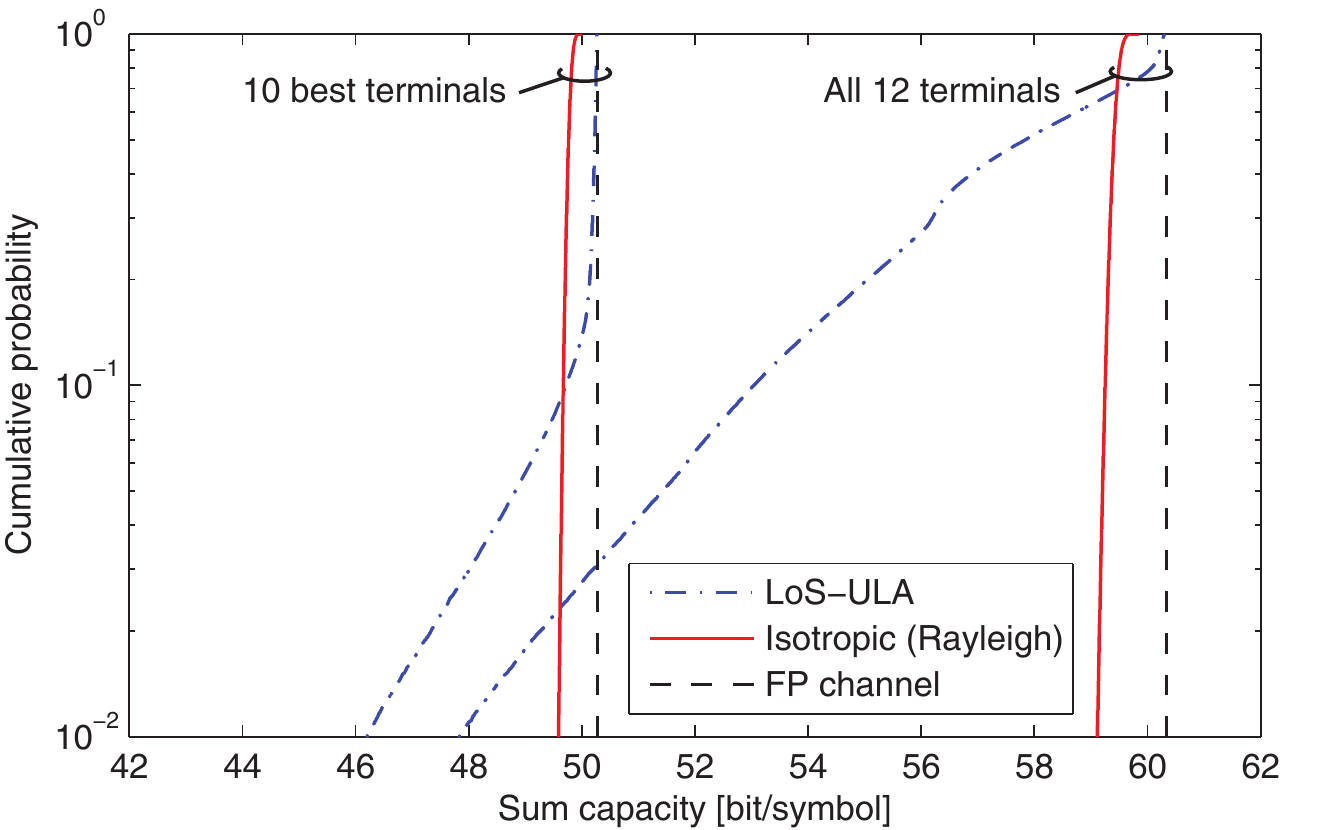} 
                \caption{}
                \label{fig:myth1} \vspace{3mm}
        \end{subfigure} \\ 
        \begin{subfigure}[b]{\columnwidth} \centering
                \includegraphics[width=\columnwidth]{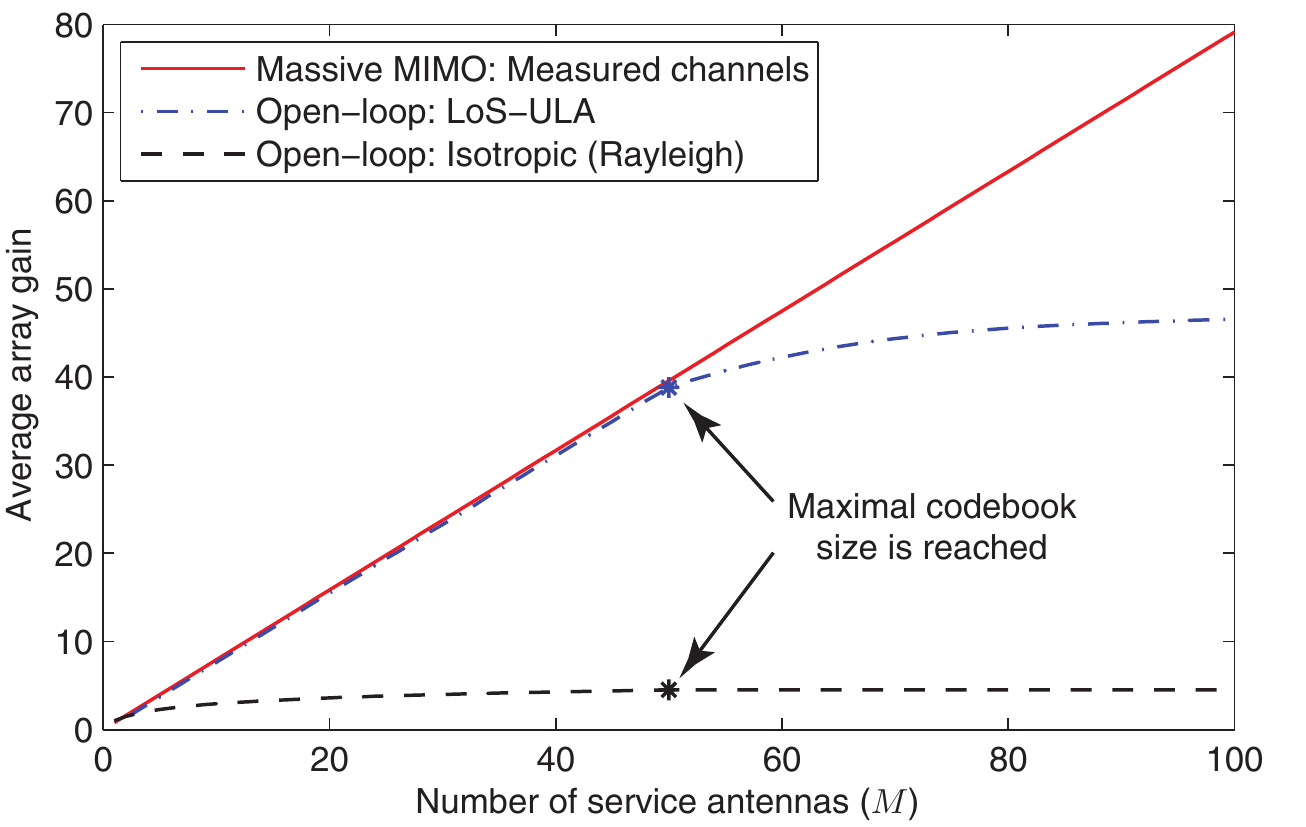} 
                \caption{}
                \label{fig:myth2}
        \end{subfigure}
        \caption{Comparison of system behavior with i.i.d.\ Rayleigh fading respectively line-of-sight propagation. There are $K = 12$ terminals and $\mathrm{SNR}_u = -5$ dB. (a) Cumulative distribution of the uplink sum capacity with $M=100$ service antennas, when either all 12 terminals or only the 10 best terminals are served. (b) Average array gain achieved for different number of service antennas. The uplink channel estimation in Massive MIMO always provides a linear slope, while the performance of open-loop beamforming depends strongly on the propagation environment and codebook size.}\label{fig:myth1and2}
\end{figure}

\textbf{Example:} Suppose the BS uses a uniform linear array (ULA) with half-wavelength antenna spacing. We compare two extreme opposite environments in Figure~\ref{fig:myth1}: (i) non-LoS isotropic scattering (i.i.d.\  Rayleigh fading) and (ii) line-of-sight (LoS) propagation. In the LoS case, the angle to each terminal determines the channel and this angle is uniformly distributed. The simulation considers $M=100$ service antennas, $K=12$ terminals, perfect CSI, and an uplink SNR of $\mathrm{SNR}_u = -5$ dB. The figure shows the cumulative probability of achieving a certain sum capacity, and the dashed vertical lines in Figure~\ref{fig:myth1} indicate the sum capacity achieved under FP.

The isotropic scattering case provides, as expected, a sum capacity close to the FP upper bound. The sum capacity in the LoS case is similar to that of isotropic scattering in the majority of cases, but there is a 10 \% risk that the LoS performance loss is more than 10 \%. The reason is that there is substantial probability that two terminals have similar angles \cite{ngo2014aspects}. A simple solution is to drop a few ``worst'' terminals from service in each coherence block; Figure~\ref{fig:myth1} illustrates this by dropping 2 out of the 12 terminals. In this case, LoS propagation offers similar performance as isotropic fading.

Since isotropic and LoS propagation represent two rather ``extreme'' environments and both are favorable for the operation of Massive MIMO, we expect that real propagation environments---which are likely to be in between these extremes---would also be favorable. This observation offers an explanation for the FP characteristics of Massive MIMO channels consistently seen in measurement campaigns (e.g., in \cite{Gao2015a}).

\subsection*{\small Myth 3: Massive MIMO performance can be achieved by open-loop beamforming techniques}

The precoding and combining in Massive MIMO rely on measured/estimated channel responses to each of the terminals and provide an array gain of $c_{\textrm{\tiny CSI}} M$ in any propagation environment \cite{ngo2014aspects}---without relying on any particular array geometry or calibration. The BS obtains estimates of the channel responses in the uplink, by receiving $K$ mutually orthogonal pilot signals transmitted by the $K$ terminals. Hence, the required pilot resources scale with $K$ but not with $M$.

By way of contrast, open-loop beamforming (OLB) is a classic technique where the BS has a codebook of $L$ pre-determined beamforming vectors and sends a downlink pilot sequence through each of them. Each terminal then reports which of the $L$ beams that has the largest gain and feeds back an index in the uplink (using $\log_2(L)$ bits). The BS transmits to each of the $K$ terminals through the beam that each terminal reported to be the best. OLB is particularly intuitive in LoS propagation scenarios, where the $L$ beamforming vectors correspond to different angles-of-departure from the array. The advantage of OLB is that no channel reciprocity or high-rate feedback is needed.  There are two serious drawbacks, however:  First, the pilot resources required are significant, because $L$ pilots are required in the downlink and $L$ should be proportional to $M$ (in order to explore and enable exploitation of all channel dimensions). Second, the $\log_2(L)$-bits-per-terminal feedback does not enable the BS to learn the channel responses accurately enough to facilitate true spatial multiplexing. This last point is illustrated by the next example.

\textbf{Example:} Figure \ref{fig:myth2} compares the array gain of Massive MIMO with that of OLB for the same two cases as in Myth 2: (i) non-LoS isotropic scattering (i.i.d.\  Rayleigh fading) and (ii) LoS propagation with a ULA. The linear array gain with MR processing is $c_{\textrm{\tiny CSI}} M$, where $c_{\textrm{\tiny CSI}} = ({1} + \frac{1}{K \cdot \mathrm{SNR}_u})^{-1}$ is the quality of the CSI (proportional to the mean-squared power of the estimate). With $K = 12$ and $\mathrm{SNR}_u = -5$ dB, the array gain is $c_{\textrm{\tiny CSI}} M \approx 0.79 M$ for Massive MIMO in both cases. For OLB, we use the codebook size of $L = M$ for $M\leq 50$ and $L = 50$ for $M>50$ in order to model a maximum permitted pilot overhead. The codebooks are adapted to each scenario, by quantizing the search space uniformly. OLB provides a linear slope in Figure \ref{fig:myth2} for $M\leq 50$ in the LoS case, but the array gain saturates when the maximum codebook size manifests itself---this would happen even earlier if the antennas are slightly misplaced in the ULA. The performance is much worse in the isotropic case, where only the logarithmic array gain $\log(M)$ is obtained before the saturation occurs. The explanation is the finite-size codebook which needs to quantize all $M$ dimensions in the isotropic case since all directions of the $M$-dimensional channel vector are equally probable. In contrast, an LoS channel direction is fully determined by the angle of arrival and thus the codebook only needs to quantize this angle.

In summary, conventional OLB provides decent array gains for small arrays in LoS propagation, but is not scalable (in terms of overhead or array tolerance) and not able to handle isotropic fading. In practice, the channel of a particular terminal might not be isotropically distributed, but have distinct statistical spatial properties. The codebook in OLB can unfortunately not be tailored to a specific terminal, but needs to explore all channel directions that are possible for the array. For large arrays with arbitrary propagation properties, the channels must be measured by pilot signaling as is done in the Massive MIMO protocol.

\subsection*{\small Myth 4: The case for Massive MIMO relies on asymptotic results}

The seminal work \cite{Marzetta2010a} on Massive MIMO studied the asymptotic regime where the number of service antennas $M \rightarrow \infty$. Numerous later works, including \cite{Ngo2013a,Hoydis2013a,Bjornson2016a}, have derived closed-form achievable spectral efficiency expressions (unit: bit/s/Hz) that are valid for any number of antennas and terminals, any SNR, and any choice of pilot signaling. These formulas do not rely on idealized assumptions such as perfect CSI, but rather on worst-case assumptions regarding the channel acquisition and signal processing. Although the total spectral efficiency per cell is greatly improved with Massive MIMO technology, the anticipated performance per user lies in the conventional range of 1-4 bit/s/Hz \cite{Bjornson2016a}---this is part of the range where off-the-shelf channel codes perform closely to the Shannon limits.

\begin{figure}[t!]
\begin{center}
\includegraphics[width=\columnwidth]{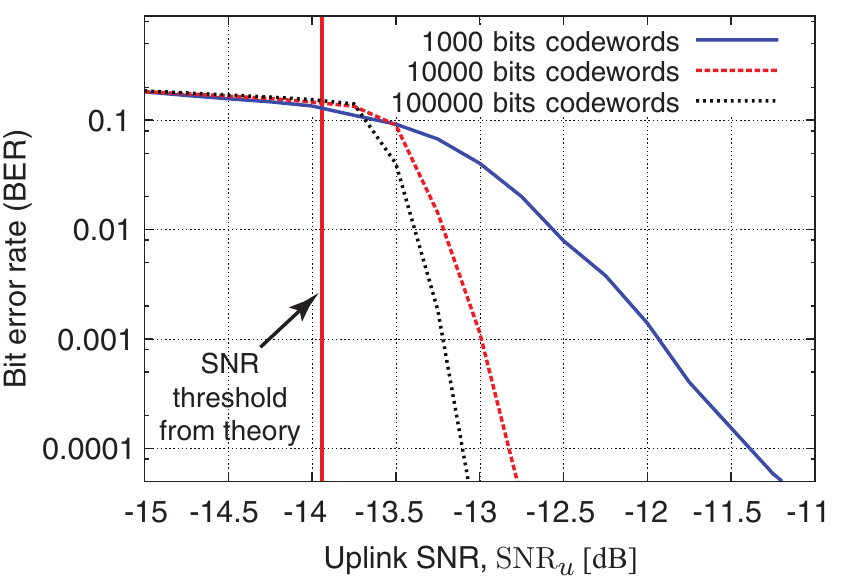}
\end{center}
\caption{Empirical uplink link performance of Massive MIMO with $M=100$ antennas and $K=30$ terminals using QPSK modulation with 1/2 coding rate and estimated channels. The vertical red line is the SNR threshold where zero BER can be achieved for infinitely long codewords, according to the spectral efficiency expression in Eq.~\eqref{eq:uplink-capacity}.} \label{fig:myth5} 
\end{figure}

\textbf{Example:} To show these properties, Figure~\ref{fig:myth5} compares the empirical link performance of a Massive MIMO system with the uplink spectral efficiency expression in Eq.~\eqref{eq:uplink-capacity}. We consider $M=100$ service antennas, $K=30$ terminals, and estimated channels using one pilot per terminal. Each terminal transmits with QPSK modulation followed by LDPC coding with rate 1/2, leading to a net spectral efficiency of 1 bit/s/Hz/terminal; that is, 30 bit/s/Hz in total for the cell. By equating Eq.~\eqref{eq:uplink-capacity} to the same target of 30 bit/s/Hz, we obtain the uplink SNR threshold $\mathrm{SNR}_u = -13.94\ \mbox{dB}$, as the value needed to achieve this spectral efficiency.

Figure~\ref{fig:myth5} shows the bit error rate (BER) performance for different lengths of the codewords, and the BER curves drop quickly as the length of the codewords increases. The vertical line indicates $\mathrm{SNR}_u = -13.94\ \mbox{dB}$, where zero BER is achievable as the codeword length goes to infinity. Performance close to this bound is achieved even at moderate codeword lengths, and part of the gap is also explained by the shaping loss of QPSK modulation and the fact that the LDPC code is optimized for AWGN channels (which is actually a good approximation in Massive MIMO due to the channel hardening). Hence, expressions such as Eq.~\eqref{eq:uplink-capacity} are well-suited to predict the performance of practical systems and useful for resource allocation tasks such as power control (see Myth 9).

\subsection*{\small Myth 5: Too much performance is lost by linear processing}

Favorable propagation, where the terminals' channels are mutually orthogonal, is a property that is generally not fully satisfied in practice; see Myth 2. Whenever there is a risk for inter-user interference, there is room for interference suppression techniques. Non-linear signal processing schemes achieve the sum capacity under perfect CSI: dirty paper coding (DPC) in the downlink and successive interference cancellation (SIC) in the uplink. DPC/SIC remove interference in the encoding/decoding step, by exploiting knowledge of what certain interfering streams will be. In contrast, linear processing can only reject interference by linear projections, for example as done with ZF. The question is how much performance that is lost by linear processing, as compared to the optimal DPC/SIC.

\begin{figure}
        \centering
        \begin{subfigure}[b]{\columnwidth} \centering 
                \includegraphics[width=\columnwidth]{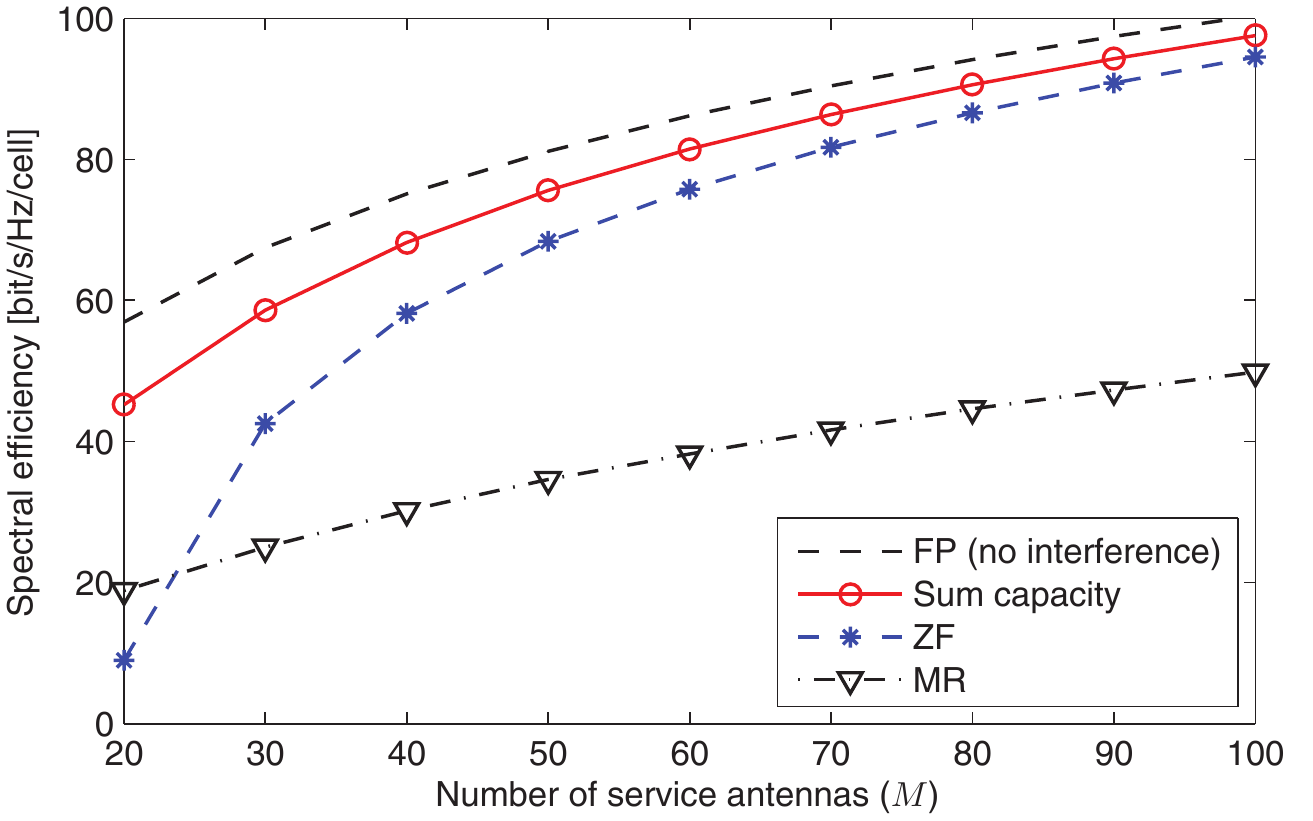} 
                \caption{}
                \label{figure_DPCvsLinear} \vspace{3mm}
        \end{subfigure} \\ 
        \begin{subfigure}[b]{\columnwidth} \centering
                \includegraphics[width=\columnwidth]{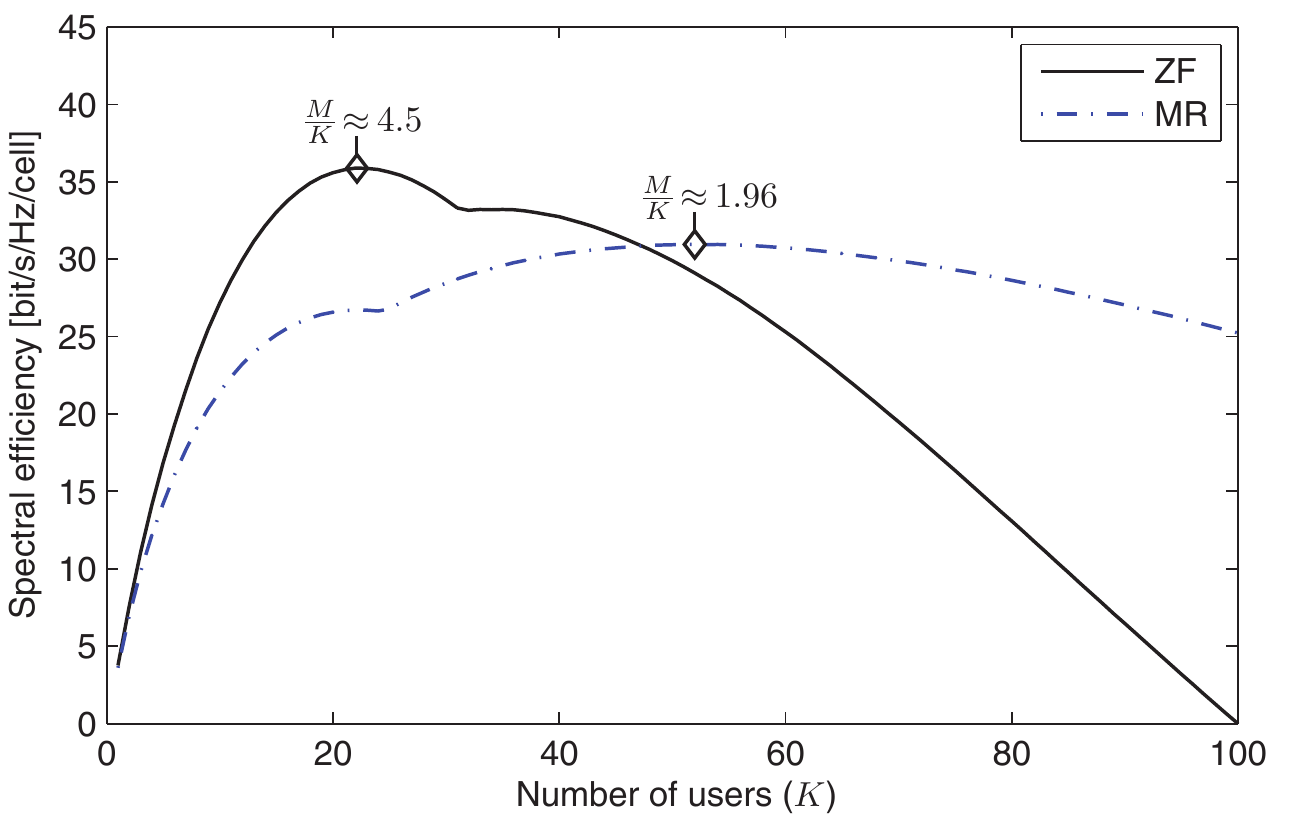} 
                \caption{}
                \label{figure_OptimalKgivenM} 
        \end{subfigure} 
        \caption{Sum spectral efficiency for i.i.d. Rayleigh fading channels with linear processing.
        (a) The sum capacity achieved by DPC/SIC is compared with linear processing, assuming perfect CSI, no inter-cell interference, and $K=20$ terminals. The loss incurred by linear processing is large when $M \approx K$, but reduces quickly as the number of antennas increases. In fact, ZF with around $M+10$ antennas gives performance equivalent to the capacity with $M$ antennas. (b) Performance in a multi-cellular system with a coherence block of $\tau = 200$ symbols, $M=100$ service antennas, estimated CSI, and an SNR of $-5$ dB. The performance is shown as a function of $K$, with ZF and MR processing. The maximum at each curve is marked and it is clear that $M/K < 10$ at these operating points.}\label{fig:impactofM}
\end{figure}

\textbf{Example:} A quantitative comparison is provided in Figure \ref{figure_DPCvsLinear} considering the sum capacity of a single cell with perfect CSI (since the capacity is otherwise unknown). The results are representative for both the uplink and downlink, due to duality. There are $K=20$ terminals and a variable number of service antennas. The channels are i.i.d.~Rayleigh fading and $\mathrm{SNR}_u = \mathrm{SNR}_d = -5$ dB.

Figure \ref{figure_DPCvsLinear} shows that there is indeed a performance gap between the capacity-achieving DPC/SIC and the suboptimal ZF, but the gap reduces quickly with $M$ since the channels decorrelate---all the curves get closer to the FP curve. Non-linear processing only provides a large gain over linear processing when $M \approx K$, while the gain is small in Massive MIMO cases with $M/K>2$. Interestingly, we can achieve the same performance as with DPC/SIC by using ZF processing with a few extra antennas (e.g., 10 antennas in this example), which is a reasonable price to pay for the much relaxed computational complexity of ZF. The gap between ZF and MR shrinks considerably when inter-cell interference is considered, as shown below.

\subsection*{\small Myth 6: Massive MIMO requires an order of magnitude more antennas than users}

For a given set of terminals, the spectral efficiency always improves by adding more service antennas---because of the larger array gain and the FP property described in Myth 2. This might be the reason why Massive MIMO is often referred to as systems with at least an order of magnitude more service antennas than terminals; that is, $M/K>10$. In general, the number of service antennas, $M$, is fixed in a deployment and not a variable, while the number of terminals, $K$, is the actual design parameter. The scheduling algorithm decides how many terminals that are admitted in a certain coherence block, with the goal of maximizing some predefined system performance metric.

\textbf{Example:} Suppose the sum spectral efficiency is the metric considered in the scheduler. Figure~\ref{figure_OptimalKgivenM} shows this metric as a function of the number of scheduled terminals, for a multi-cellular Massive MIMO deployment of the type considered in \cite{Bjornson2016a}. There are $M=100$ service antennas per cell. The results are applicable in both the uplink and the downlink, if power control is applied to provide an SNR of $-5$ dB for every terminal. A relatively short coherence block of $\tau = 200$ symbols is considered and the pilot reuse across cells is optimized (this is why the curves are not smooth). The operating points that maximize the performance, for ZF and MR processing, are marked and the corresponding values of the ratio $M/K$ are indicated. Interestingly, the optimized operating points are all in the range $M/K < 10$; thus, it is not only possible to let $M$ and $K$ be at the same order of magnitude, it can even be desirable. With MR processing, the considered Massive MIMO system operates efficiently also at $M=K=100$ which gives $M/K=1$; the rate per terminal is small at this operating point but the sum spectral efficiency is not. We also stress that there is a wide range of $K$-values that provides almost the same sum performance, showing the ability to share the throughput between many or few terminals by scheduling.

In summary, there are no strict requirements on the relation between $M$ and $K$ in Massive MIMO. If one would like to give a simple definition of a Massive MIMO setup, it is a system with unconventionally many active antenna elements, $M$, that can serve an unconventionally large number of terminals, $K$. One should avoid specifying a certain ratio $M/K$, since it depends on a variety of conditions; for example, the system performance metric, propagation environment, and coherence block length.

\subsection*{\small Myth 7: A new terminal cannot join the system since there is no initial array gain}

The coherent processing in Massive MIMO improves the effective SNR by a factor $c_{\textrm{\tiny CSI}} M$, where $0 < c_{\textrm{\tiny CSI}} \leq 1$ is the CSI quality (see Myth 3 for details). This array gain enables the system to operate at lower SNRs than contemporary systems. As seen from the factor $c_{\textrm{\tiny CSI}}$, the BS needs to estimate the current channel response, based on uplink pilots, to capitalize on the array gain. When a previously inactive terminal wishes to send or request data it can therefore pick one of the unused pilot sequences and contact the BS using that pilot. The system can, for example, be implemented by reserving a few pilots for random access, while all active terminals use other pilots to avoid collisions. It is less clear how the BS should act when contacting a terminal that is currently inactive---it cannot exploit any array gain since this terminal has not sent a pilot.

This question was considered in \cite{Karlsson2014a} and the solution is quite straightforward to implement. Instead of sending precoded downlink signals to the $K$ terminals, the BS can occasionally utilize the same combined transmit power to only broadcast control information within the cell (e.g., to contact inactive terminals). Due to the lack of array gain, this broadcast signal will be $c_{\textrm{\tiny CSI}} M/K$ times weaker than the user-specific precoded signals. We recall that $M/K < 10$ at many operating points of practical interest, which was noted in Myth 6 and exemplified in Figure~\ref{figure_OptimalKgivenM}. The ``loss'' $c_{\textrm{\tiny CSI}} M/K$ in effective SNR is partially compensated by the fact that the control signals are not exposed to intra-cell interference, while further improvements in reliability can be achieved using stronger channel codes. Since there is no channel hardening we can also use classical diversity schemes, such as space-time codes and coding over subcarriers, to mitigate small-scale fading.

In summary, control signals can be transmitted also from large arrays without the need for an array gain. The numerical examples in \cite{Karlsson2014a} show that the control data rate is comparable to the individual precoded payload data rates at typical operating points (due to the lack of intra-cell interference and the concentration of transmit power), but the multiplexing gain is lost since one signal is broadcasted instead of precoded transmission of $K$ separate signals.

\subsection*{\small Myth 8: Massive MIMO requires high precision hardware}

A main feature of Massive MIMO is the coherent processing over the $M$ service antennas, using measured channel responses. Each desired signal is amplified by adding the $M$ signal components coherently, while uncorrelated undesired signals are not amplified since their components add up noncoherently. 

Receiver noise and data signals associated with other terminals are two prime examples of undesired additive quantities that are mitigated by the coherent processing. There is also a third important category: distortions caused by impairments in the transceiver hardware. There are numerous impairments in practical transceivers; for example, non-linearities in amplifiers, phase noise in local oscillators, quantization errors in analog-to-digital converters, I/Q imbalances in mixers, and non-ideal analog filters. The combined effect of these impairments can be described either stochastically \cite{Bjornson2014a} or by hardware-specific deterministic models \cite{Gustavsson2014a}. In any case, most hardware impairments result in additive distortions that are substantially uncorrelated with the desired signal, plus a power loss and phase-rotation of the desired signals. The additive distortion noise caused at the BS has been shown to vanish with the number of antennas \cite{Bjornson2014a}, just as conventional noise and interference, while the phase-rotations from phase noise remain but are not more harmful to Massive MIMO than to contemporary systems. We refer to \cite{Bjornson2014a} and \cite{Gustavsson2014a} for numerical examples that illustrate these facts.

In summary, the Massive MIMO gains are not requiring high-precision hardware; in fact, lower hardware precision can be handled than in contemporary systems since additive distortions are suppressed in the processing. Another reason for the robustness is that Massive MIMO can achieve extraordinary spectral efficiencies by transmitting low-order modulations to a multitude of terminals, while contemporary systems require high-precision hardware to support high-order modulations to a few terminals.

\subsection*{\small Myth 9: With so many antennas, resource allocation and power control is hugely complicated}

Resource allocation usually means that the time-frequency resources are divided between the terminals, to satisfy user-specific performance constraints, to find the best subcarriers for each terminal, and to combat the small-scale fading by power control. Frequency-selective resource allocation can bring substantial improvements when there are large variations in channel quality over the subcarriers, but it is also demanding in terms of channel estimation and computational overhead since the decisions depend on the small-scale fading that varies at the order of milliseconds. If the same resource allocation concepts would be applied in Massive MIMO systems, with tens of terminals at each of the thousands of subcarriers, the complexity would be huge.

Fortunately, the channel hardening effect in Massive MIMO means that the channel variations are negligible over the frequency domain and mainly depend on large-scale fading in the time domain, which typically varies 100--1000 times slower than the small-scale fading. This renders the conventional resource allocation concepts unnecessary. The whole spectrum can be simultaneously allocated to each active terminal and the power control decisions are made jointly for all subcarriers based only on the large-scale fading characteristics.

\textbf{Example:} Suppose we want to provide uniformly good performance to the terminals in the downlink. This resource allocation problem is only non-trivial when the $K$ terminals have different average channel conditions. Hence, we associate the $k$th terminal with a user-specific CSI quality $c_{\textrm{\tiny CSI},k}$, a nominal downlink SNR value of $\mathrm{SNR}_{d,k}$ when the transmit power is shared equally over the terminals, and a power-control coefficient $\eta_k \in [0,K]$ that is used to reallocate the power over the terminals (under the constraint $\sum_{k=1}^{K} \eta_k \leq K$). By generalizing the spectral efficiency expression in Eq.~\eqref{eq:uplink-capacity} to cover these user-specific properties (and dropping the constant pre-log factor), we arrive at the following optimization problem:
\begin{align} \label{eq:power-allocation-original}
\maximize{ \substack{ \eta_1,\ldots,\eta_K \in [0,K]  \\ \sum_{k}\eta_k \leq K} } &\,\, \min_{k} \quad \log_2 \left( 1+ \frac{ c_{\textrm{\tiny CSI},k} \cdot M  \cdot \mathrm{SNR}_{d,k} \cdot \eta_k}{\mathrm{SNR}_{d,k} \sum_{i=1}^{K} \eta_i + 1} \right) 
\\ &\Updownarrow \notag \\  \notag
\maximize{\substack{ \eta_1,\ldots,\eta_K \in [0,K]  \\ \sum_{k}\eta_k \leq K, \, R \geq 0 }  } &\quad R  \\ \textrm{subject to} \,\,\,\,\,\,\,\, & \quad
 c_{\textrm{\tiny CSI},k} \cdot M \cdot \mathrm{SNR}_{d,k} \cdot \eta_k \geq \left( 2^R - 1 \right) \\ & \quad \cdot\left( \mathrm{SNR}_{d,k} \sum_{i=1}^{K} \eta_i + 1 \right) \quad \textrm{for} \,\, k = 1,\ldots,K.
\notag
\end{align}
This resource allocation problem is known as max-min fairness, and since we maximize the worst-terminal performance the solution gives the same performance to all terminals. The second formulation in Eq.~\eqref{eq:power-allocation-original} is the epigraph form of the original formulation. From this reformulation it is clear that all the constraints are linear functions of the power-control coefficients $\eta_1,\ldots,\eta_K$, thus Eq.~\eqref{eq:power-allocation-original} is a linear optimization problem for every fixed worst-terminal performance $R$. The whole problem is solved by line search over $R$, to find the largest $R$ for which the constraints are feasible. In other words, the power control optimization is a so-called quasi-linear problem and can be solved by standard techniques (e.g., interior point methods) with low computational complexity. We stress that the power control in Eq.~\eqref{eq:power-allocation-original} only depends on the large-scale fading---the same power control can be applied on all subcarriers and over a relatively long time period. 

To summarize, the resource allocation can be greatly simplified in Massive MIMO systems. It basically reduces to admission control (which terminals should be active) and long-term power control (in many cases it is a quasi-linear problem). The admitted terminals may use the full bandwidth---there is no need for frequency-selective allocation when there is no frequency-selective fading. The complexity of power control problems such as Eq.~\eqref{eq:power-allocation-original} scales with the number of terminals, but is independent of the number of antennas and subcarriers.

\subsection*{\small Myth 10: With so many antennas, the signal processing complexity will be overwhelming}

\begin{figure*}
  \includegraphics[width=\textwidth]{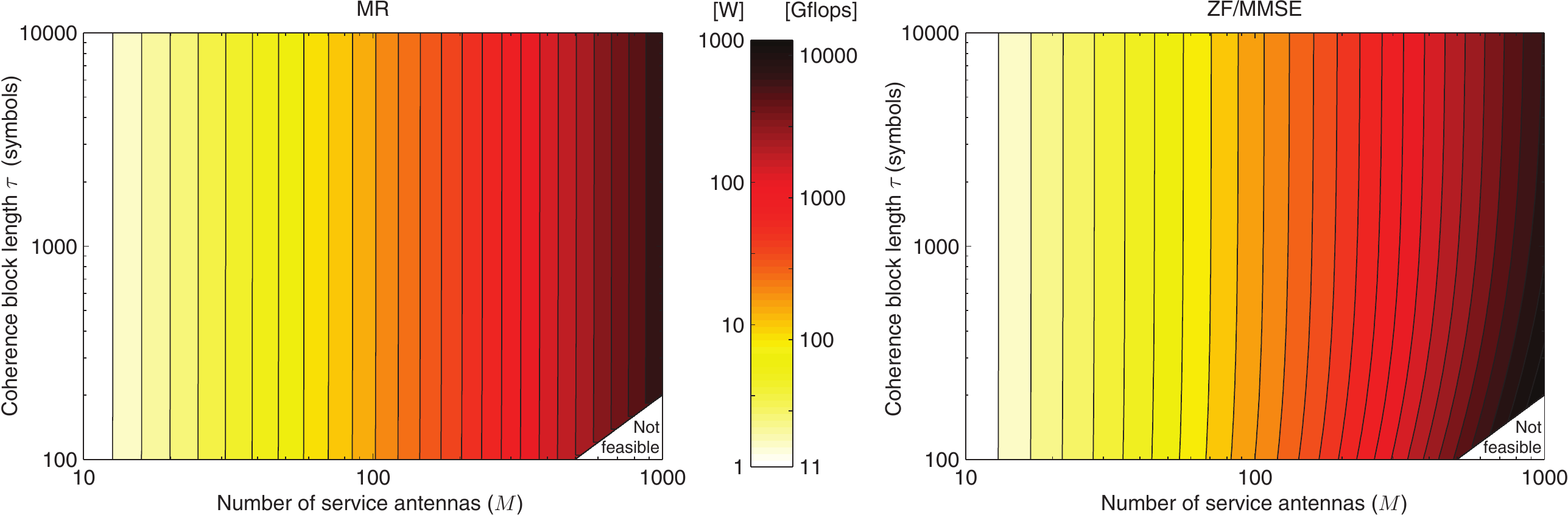} 
  \caption{Computational complexity (in flops) of the main baseband signal processing operations in an OFDM Massive MIMO setup: FFTs, channel estimation, precoding/combining of payload data, and computation of precoding/combining matrices. The complexity is also converted into an equivalent power consumption using a typical computational efficiency of $12.8$ Gflops/Watt \cite{Schneider2012a}.} \label{figure_Complexity}
\end{figure*}

The baseband processing is naturally more computationally demanding when having $M > 1$ BS antennas that serves $K>1$ terminals, as compared to only serving one terminal using one antenna port. The important question is how fast the complexity increases with $M$ and $K$; is the complexity of a typical Massive MIMO setup manageable using contemporary or future hardware generations, or is it totally off the charts?

In an OFDM implementation of Massive MIMO, the signal processing needs to take care of a number of tasks; for example, fast Fourier transform (FFT), channel estimation using uplink pilots, precoding/combining of each payload data symbol (a matrix-vector multiplication), and computation of the precoding/combining matrices. The complexity of these signal processing tasks scales linearly with the number of service antennas, and everything except the FFT complexity also increases with the number of terminals. The computation of a precoding/combining matrix depends on the processing scheme: MR has a linear scaling with $K$, while ZF/MMSE have a faster scaling since these involve matrix inversions.
Nevertheless, all of these processing tasks are standard operations for which the required number of floating point operations per second (flops) are straightforward to compute \cite{Yang2013b}. This can provide rough estimates of the true complexity, which also depends strongly on the implementation and hardware characteristics.

\textbf{Example:} To exemplify the typical complexity, suppose we have a 20 MHz bandwidth, $1200$ OFDM subcarriers, and an oversampling factor of 1.7 in the FFTs. Figure \ref{figure_Complexity} shows how the computational complexity depends on the length $\tau$ of the coherence block and on the number of service antennas $M$. The number of terminals are taken as $K = M/5$, which was a reasonable ratio according to Figure \ref{figure_OptimalKgivenM}. Results are given for both MR and ZF/MMSE processing at the BS. Each color in Figure \ref{figure_Complexity} represents a certain complexity interval, and the corresponding colored area shows the operating points that give a complexity in this interval. The complexities can also be mapped into a corresponding power consumption; to this end, we consider the state-of-the-art DSP in \cite{Schneider2012a}  which has a computational efficiency of $E = 12.8$ Gflops/Watt.

Increasing the coherence block means that the precoding/combining matrices are computed less frequently, which reduces the computational complexity. This gain is barely visible for MR, but can be substantial for ZF/MMSE when there are very many antennas and terminals (since the complexity of the matrix inversion is then large). For the typical operating point of $M=200$ antennas, $K=40$ terminals, and $\tau = 200$ symbols, the complexity is 559 Gflops with MR and 646 Gflops with ZF/MMSE. This corresponds to 43.7 Watt and 50.5 Watt, respectively, using the exemplified DSP. These are feasible complexity numbers even with contemporary technology; in particular, because the majority of the computations can be parallelized and distributed over the antennas. It is only the computation of the precoding/combining matrices and the power control that may require a centralized implementation.

In summary, the baseband complexity of Massive MIMO is well within the practical realm. The complexity difference between MR and ZF/MMSE is relatively small since the precoding/combining matrices are only computed once per coherence block---the bulk of the complexity comes from FFTs and matrix-vector multiplications performed at a per symbol basis.

\section*{\small The Critical Question}

\subsection*{\small Can Massive MIMO work in FDD operation?}

\begin{figure*}[t!]
\begin{center}
\includegraphics[width=1.7\columnwidth]{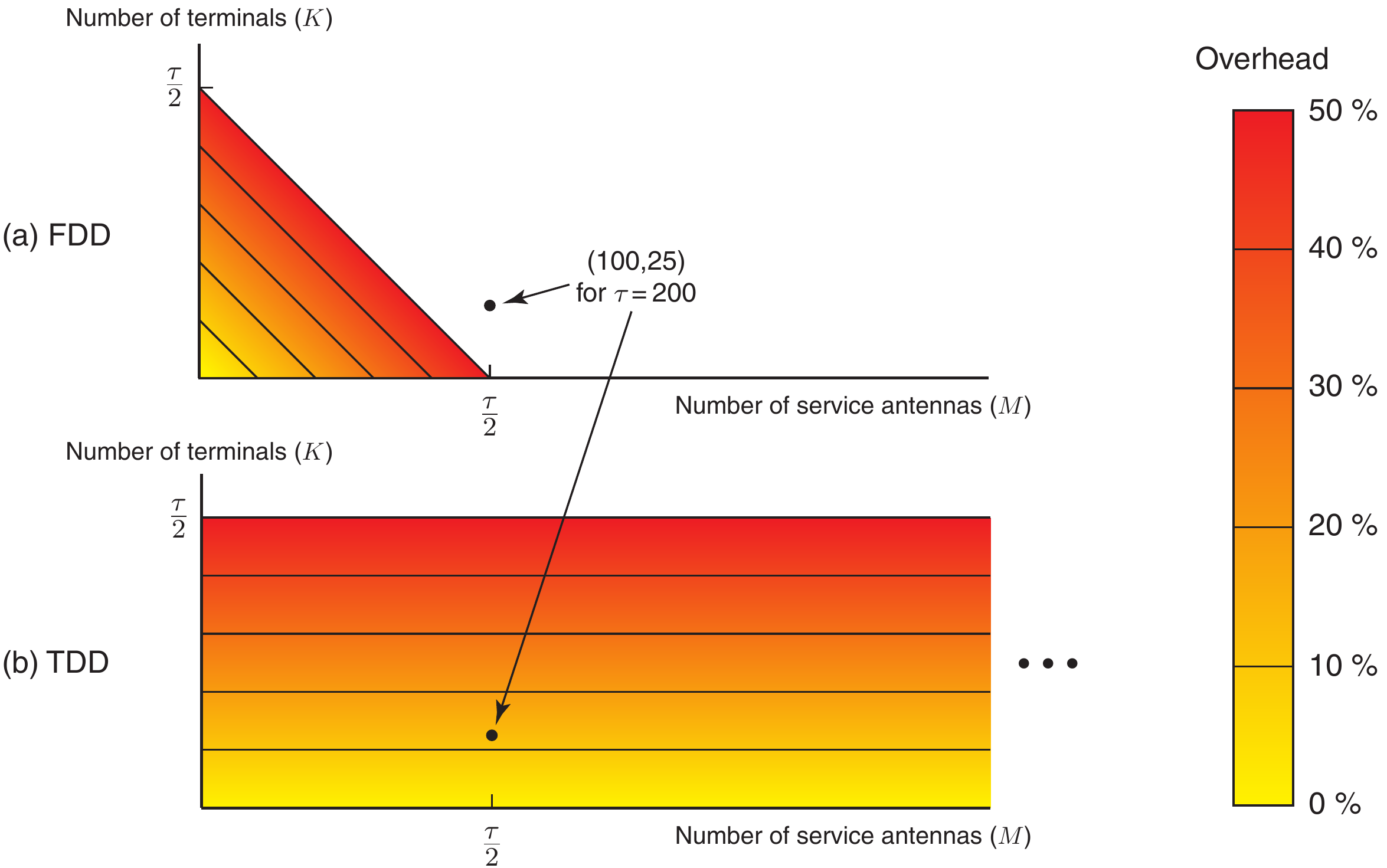}
\end{center}
\caption{Illustration of the typical overhead signaling in Massive MIMO based on (a) FDD and (b) TDD operation. The main difference is that FDD limits the number of antennas, while TDD can have any number of antennas. For a coherence block with $\tau=200$ even a modest Massive MIMO setup with $M=100$ and $K=25$ is only supported in TDD operation.} \label{fig:question} 
\end{figure*}

The canonical Massive MIMO protocol, illustrated in Figure \ref{figure_uplink-downlink}, relies on TDD operation. This is because the BS processing requires CSI and the overhead of CSI acquisition can be greatly reduced by exploiting channel reciprocity. Many contemporary networks are, however, operating in frequency-division duplex (FDD) mode, where the uplink and downlink use different frequency bands and channel reciprocity cannot be harnessed. The adoption of Massive MIMO technology would be much faster if the concept could be adapted to also operate in FDD. But the critical question is: Can Massive MIMO work in FDD operation?

To explain the difference between TDD and FDD, we describe the related CSI acquisition overhead. Recall that the length of a coherence block is $\tau = B_c T_c$ symbols. Massive MIMO in TDD mode uses $K$ uplink pilot symbols per coherence block and the channel hardening eliminates the need for downlink pilots. In contrast, a basic FDD scheme requires $M$ pilot symbols per coherence block in the downlink band, and $K$ pilot symbols plus feedback of $M$ channel coefficients per terminal on the uplink band (e.g., based on analog feedback using $M$ symbols and multiplexing of $K$ coefficients per symbol). Hence, it is the $M+K$ uplink symbols per coherence block that is the limiting factor in FDD. The feasible operating points $(M,K)$ with TDD and FDD operations are illustrated in Figure \ref{fig:question}, as a function of $\tau$, and are colored based on the percentage of overhead that is needed.

The main message from Figure \ref{fig:question} is that TDD operation supports any number of service antennas, while there is a tradeoff between antennas and terminals in FDD operation. The extra FDD overhead might be of little importance when $\tau = 5000$ (e.g., in low mobility scenarios at low frequencies), but it is a critical limitation when $\tau = 200$ (e.g., for high mobility scenarios or at higher frequencies). For instance, the modest operating point of $M=100$ and $K=25$ is marked in Figure \ref{fig:question} for the case of $\tau = 200$. We recall that this was a good operating point in Figure~\ref{figure_OptimalKgivenM}. This point can be achieved with only 12.5 \% pilot overhead in TDD operation while FDD cannot even support it by spending 50 \% of the resources on overhead signaling. It thus appears that FDD can only support Massive MIMO in special low-mobility and low-frequency scenarios.

Motivated by the demanding CSI acquisition in FDD mode, several research groups have proposed methods to reduce the overhead; two excellent examples are \cite{Yin2013a} and \cite{Adhikary2014a}. Generally speaking, these methods assume that there is some kind of channel sparsity that can be utilized; for example, a strong spatial correlation where only a few strong eigendirections need to be estimated or that the impulse responses are sparse in time. While methods of these kinds achieve their goals, we stress that the underlying sparsity assumptions are so far only hypotheses. Measurements results available in the literature indicate that spatial sparsity assumptions are questionable at lower frequencies (see e.g., Figure~4 in \cite{Gao2015a}). At millimeter wave frequencies, however, the channel responses may indeed be sparse \cite{Adhikary2014a}.

The research efforts on Massive MIMO in recent years have established many of the key characteristics of the technology, but it is still unclear to what extent Massive MIMO can be applied in FDD mode. We encourage researchers to investigate this thoroughly in the coming years, to determine if any of the sparsity hypotheses are indeed true or if there are some other ways to reduce the overhead signaling. Proper answers to these questions require intensive research activities and channel measurements.

\section*{Acknowledgment}

We would like to thank all of our Massive MIMO research collaborators; in particular, Prof.~Fredrik Tufvesson and his colleagues at Lund University who provided the picture of their LuMaMi testbed. The writing of this article was supported by the EU FP7 under ICT-619086 (MAMMOET) and by ELLIIT and CENIIT.

\section*{Authors}

{\bf\em Emil Bj\"ornson} (emil.bjornson@liu.se) received the Ph.D. degree in 2011 from KTH Royal Institute of Technology, Sweden. He was a joint postdoctoral researcher at Sup\'{e}lec, France, and at KTH Royal Institute of Technology, Sweden. Since 2014 he is a Research Fellow at Link\"{o}ping University, Sweden. He is the first author of the textbook \emph{Optimal Resource Allocation in Coordinated Multi-Cell Systems} and he received the 2014 Outstanding Young Researcher Award from IEEE ComSoc EMEA, the 2015 Ingvar Carlsson Award, and best conference paper awards in 2009, 2011, 2014, and 2015.

{\bf\em Erik G. Larsson} (erik.g.larsson@liu.se) is Professor at Link\"{o}ping University in Sweden. He has been Associate Editor for several IEEE journals, he serves as chair of the IEEE SPS SPCOM technical committee in 2015, as chair of the steering committee for the IEEE Wireless Communications Letters in 2014?2015, and as General Chair of the Asilomar SSC Conference 2015. He received the IEEE Signal Processing Magazine Best Column Award twice, in 2012 and 2014 and the IEEE ComSoc Stephen O. Rice Prize in Communications Theory in 2015.

{\bf\em Thomas L. Marzetta} (tom.marzetta@alcatel-lucent.com) is the originator of Massive MIMO, the most promising technology available to address the ever increasing demand for wireless throughput. He is a Bell Labs Fellow and Group Leader of Large Scale Antenna Systems at Bell Labs, Alcatel-Lucent and Co-Head of their FutureX Massive MIMO project. Dr.~Marzetta  received the Ph.D. and S.B. in Electrical Engineering from Massachusetts Institute of Technology in 1978 and 1972, and the M.S. in Systems Engineering from University of Pennsylvania in 1973.
Currently, Dr.~Marzetta serves as Coordinator of the GreenTouch Consortium's Large-Scale Antenna Systems Project and as Member of the Advisory Board of MAMMOET (Massive MIMO for Efficient Transmission), an EU-sponsored FP7 project. For his achievements in Massive MIMO he has received the 2015 IEEE W.~R.~G.~Baker Award, the 2014 Thomas Alva Edison Patent Award in Telecommunications category, the 2014 GreenTouch 1000x Award, the 2013 IEEE Guglielmo Marconi Prize Paper Award, and the 2013 IEEE OnlineGreenComm Conference Best Paper Award. He was elected a Fellow of the IEEE in 2003.

\bibliographystyle{IEEEbib}
\bibliography{IEEEabrv,refs}

\end{document}